# SPEAKER RECOGNITION WITH A MLP CLASSIFIER AND LPCC CODEBOOK[1]

*Daniel Rodriguez-Porcheron, Marcos Faundez-Zanuy*

Escola Universitària Politècnica de Mataró
Universitat Politècnica de Catalunya(UPC)
Avda. Puig i Cadafalch 101-111, E-08303 MATARO (BARCELONA)
e-mail: faundez@eupmt.upc.es  http://www.eupmt.upc.es/veu

## ABSTRACT

This paper improves the speaker recognition rates of a MLP classifier and LPCC codebook alone, using a linear combination between both methods. In our simulations we have obtained an improvement of 4.7% over a LPCC codebook of 32 vectors and 1.5% for a codebook of 128 vectors (error rate drops from 3.68% to 2.1%). Also we propose an efficient algorithm that reduces the computational complexity of the LPCC-VQ system by a factor of 4.

## 1. INTRODUCTION

Several methods exist for speaker recognition purposes. The most successful are: VQ [1], HMM [2] and GMM [3]. Also the neural nets can achieve good results. During last years several methods have been proposed, such as Time Delay Neural Networks(TDNNs) [4], Radial Basis Function (RBF) [5], Learning Vector Quantization (LVQ) [6], Self Organizing Maps (SOM) [7], etc.

Other methods that achieve poor results can be improved in several ways. In [8] Farrell evaluates a MLP classifier and he obtains that it can be improved using a tree of Single Layer Perceptron (SLP) classifiers.

In this paper we propose a combination of a VQ speaker recognition system and a MLP classifier that improves the results of both systems alone. The problem of computational burden is also addressed, proposing a new structure (Figure 1).

Our work is centered on speaker recognition in text independent mode.

## 2. SPEAKER RECOGNITION USING VQ AND MLP

### 2.1 Database

Our experiments have been computed over 38 speakers from the New England dialect of the DARPA TIMIT Database (24 males&14 females). The speech samples were downsampled from 16 kHZ to 8kHZ, and pre-emphasized by a first order filter whose transfer function was $H(z)=1-0.95z^{-1}$. A 30 ms Hamming window was used, and the overlapping between adjacent frames was 2/3. A cepstral vector of order 12 was computed from the LPC coefficients. Five sentences are used for training, and 5 sentences for testing (each sentence is between 0.9 and 2.8 seconds long).

### 2.2 Speaker recognition using LPCC-VQ

In this system each speaker is modeled with a vector quantizer during the training process. The identification is done quantizing the input sentence with all the codebooks and choosing the quantizer that gives the lowest accumulated error. A detailed explanation of this system can be found in [1].

In our simulation the codebooks are generated with the LBG algorithm.[9]

### 2.3 Speaker recognition using a MLP classifier.

We have chosen the same structure than Farrell [8], that is: each speaker is modeled with a MLP of 12 input neurons (one neuron for each cepstral coefficient), one hidden layer with 16 neurons, and one output neuron. Experimentally it has been found that there is no improvement increasing the number of hidden neurons. This is because it is more complicated to obtain an accurate division of the 12 dimensional space, and thus the probability of to get stack in a local minima increases.

The MLP is trained in the following way: Using a training set of vectors, they are labeled with "1" if they belong to the speaker that it is going to be modeled, and with "0" if it is from a different speaker. Thus, MLP is trained for a "1" output when the input belongs to the speaker and a "0" output when the input is from a different speaker. In our system the MLP is trained with the Levenberg-Marquardt algorithm [10]. Obviously in a real performance, the outputs will not be "0" or "1" but they will tend to this value.

Obviously the number of inhibitory vectors ("0") is greater than the number of excitatory vectors. Hence, the classifier tends to learn that "everything" is inhibitory. We have alleviated this problem compressing the inhibitory vectors by means of a vector quantization of the inhibitory vectors.

---
[1] This work has been supported by the CICYT TIC97-1001-C02-02

In the test phase the input sentence is partitioned into frames, and the LPCC vector of each frame is computed. This vector is presented to the MLP classifier, and the output of this vector is accumulated to the output of the other vectors. This process is repeated for all the speakers models, and the speaker that gives the greatest accumulated output is selected.

An important point is the way in which the MLP is trained. We have chosen the Levenberg-Marquardt algorithm. The neural net weights are initialized with a multi-start algorithm, which consists in training 4 random initializations. Each initialization is trained during 8 epochs, and the random initialization that gives the lowest error is selected.

## 3. RESULTS

Previously to the description of our novel proposed method, we will obtain the results of the combined methods alone.

### 3.1 LPCC-VQ.

Using the method described in section 2.2 we obtain the results summarized in table 1. These results compare favorably with the results obtained by Farrell [8] (see table 1), because we use 38 speakers while he uses only 20. This difference is due to a more accurate codebook generation process (we have used the splitting algorithm, with an hyperplane method for splitting the centroids).

One important subject for obtaining good recognition rates is the codebook initialization and generation processes, empty cells processing, etc.

| Codebook size (bits) | LPCC-VQ | LPCC-VQ Farrell [8] |
|---|---|---|
| 4 | 10% | 10% |
| 5 | 6,84% | 8% |
| 6 | 6,31% | 5% |
| 7 | 3,68% | 4% |

**Table 1** Recognition errors for VQ system.

### 3.2 MLP classifier

Using the method described in section 2.3 we obtain an identification error about 20%, that obviously is higher than the results of the LPCC-VQ method of section 3.1.

We have evaluated that the correlation coefficient between both kinds of measures is about –0.7 (the LPCC-VQ produces an error measure, and the MLP produces a similitude measure. Thus it is a negative value).

### 3.3 Linear Combination between LPCC-VQ and MLP classifier.

We propose the combined scheme of figure 1 for improving the results of sections 3.1 and 3.2. The algorithm is the following:

1. To compute the cepstral vectors (LPCC) of the test input sentence.
2. These vectors are quantized with the codebooks of all the speakers, obtaining N distortion measures, where N is the number of speakers.
3. These error measures are sorted, and the K smallest values are chosen.
4. The LPCC vectors of the input frame are filtered with the K MLP classifiers assigned to the K selected speakers in step 3.
5. The K distortion measures of step 3 are combined with the K similitude measures of step 4 for obtaining one value for each of the K speakers.

The measures of similitude and distortion (error) are combined with the following expression:

$$\text{measure}_{comb} = \text{measure}_{error} - \alpha \, \text{measure}_{similitude} \quad (1)$$

where $\alpha$ is the weighting factor and it is set experimentally as described in section 3.4.

The interpretation of both measures is opposite, so one way of combining them is substracting the similitude measure to the error measure. Thus, the combined measure is an error measure.

We have tested two different distortion measures:
a) Mean square error of the LPCC vectors (MSE).
b) Mean Absolute difference of the LPCC vectors (MAD).

Table 2 summarizes the results for K=2. From table 2 it can be deducted that the linear combination improves the recognition rates of both systems alone. Main conclusions are:

- For a 4 bits codebook size there is a reduction in the error recognition rate of 4,2% with the combined scheme.
- The improvement is less significative if the codebook size is increased.
- Better results are obtained with the MAD criterion than with the MSE criterion.
- With the MAD criterion the error recognition rate drops 4% for 5 and 6 bits codebook and 1.5% for 7 bit codebook.
- With the MAD criterion the error recognition rate saturates at 2.1% for codebooks of more than 5 bits.
- The combined scheme with a codebook of 5 bits outperforms the LPCC-VQ system with a codebook of 7 bits, with a computational burden that is approximately 4 times lower, as described in the following section.

### 3.4 Relevance of $\alpha$ and K

Figure 2 shows the recognition rates as function of K for a codebook of 5 bits, in the combined scheme of figure 1. The upper line corresponds to a MLP classifier without combination. That is, the LPCC-VQ is used only as a preclassifier of the K most probable speakers, and only this K MLP classifiers are evaluated. This curve presents the following special cases:

- For K=1 the system of the bottom curve is equivalent to the LPCC-VQ system. Thus, The MLP classifier has no relevance.
- For K=N the LPCC-VQ has no relevance, and the system is equivalent to the MLP classifier of section 3.2

The bottom line corresponds to the combined scheme of figure 1, and it can be seen that for K=2 the minimal recognition error of 2.1% is achieved. This value implies the lower computational complexity (see section 3.5), so K=2 is selected in our scheme.

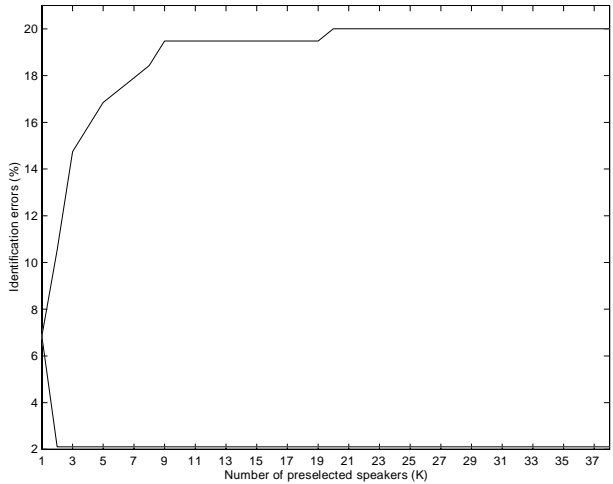

**Figure 2.** Identification errors as function of K.

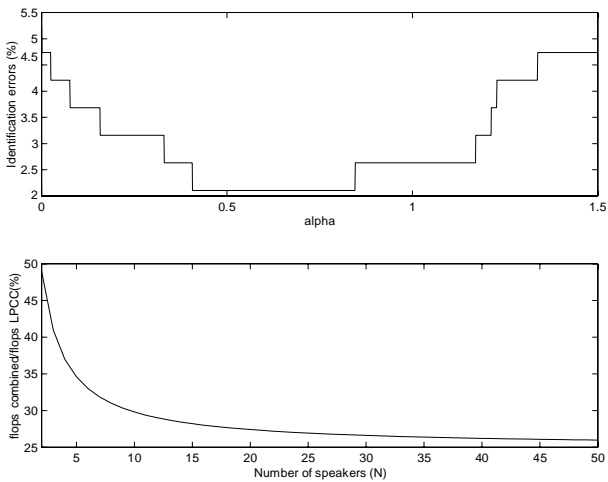

**Figure 3.** Identification errors as function of α for 5 bit codebook and number of operations of the combined system relative to the LPCC-VQ.

About the α selection we have done an exhaustive search in a wide enough range of values. Figure 3 on the top shows the evolution of the recognition errors as function of the α value. This curve has been obtained for a 5 bit codebook. Our simulations have proven that the selection of α is not a critical question because its optimal value is always around the same value.

### 3.7 Computational burden comparison.

This section compares the computational complexity of the LPCC-VQ scheme and the proposed scheme of figure 1. The number of operations needed for computing the LPCC will be neglected because it must be done in both methods. The notation is the following:

> N : number of speakers.
> P : predictive analysis order.
> $T_{c1}$ : Size of the LPCC codebook.
> K : number of preselected speakers, using the LPCC codebook
> $n_i$ : Number of neurons in the input layer.
> $n_{h1}$, $n_{h2}$ : Number of neurons in the hidden layers,
> $C_{tg}$ : Number of instructions for the nonlinear transfer function computation.

In this situation the computational burden of the LPCC-VQ is proportional to:

$$N_{LPCC} = T_{c1} \times p \times N \quad (2)$$

while the computational burden of the combined scheme is proportional to:

$$N_{comb} = N_{LPCC} + K \times \left( n_i \times n_{h1} + n_{h1} + C_{tg} \times n_{h1} \right) \quad (3)$$

With a proper substitution and for K=2 we obtain the lowest computational burden of the combined system:

$$N_{LPCC} = 1536N \quad (4)$$

$$N_{comb} = 384N + 736 \quad (5)$$

Figure 3 on the bottom shows the ratio between the number of operations of the combined system over the LPCC-VQ system.. From this figure it can be deducted that if the number of speakers in the database is greater than 20 the computational burden is approximately 4 times lower.

### 4. Conclusions

In this paper we have proposed a combined scheme using two different speaker recognition methods that improves the recognition accuracy, and reduces the computational complexity of the testing process of a conventional vector quantization system by a factor of 4.

### 7. References

[1] F. K. Soong, A. E. Rosenberg, L. R. Rabiner y B. H. Juang " A vector quantization approach to speaker recognition". ICASSP pp. 387-390, 1985

[2] M. Savic & S. K. Gupta "Variable parameter speaker verification system based on hidden Markov modeling". ICASSP pp. 281-284, 1990.

___

[3] R. C. Rose & D. A. Reynolds "Text independent speaker identification using automatic acoustic segmentation". ICASSP, pp. 293-296. 1990

[4] Y. Bennani, P. Gallinari, "On the use of TDNN-extracted features informations in talker identification", Proceedings ICASSP, pp. 385-388, 1991.

[5] J. Oglesby, J. S. Mason, "Radial basis function networks for speaker recognition", Proceedings ICASSP, pp. 393-396, 1991.

[6] Y. Bennani, P. Gallinari, "A connectionist approach for speaker identification", Proceedings ICASSP, pp. 265-268, 1990.

[7] E. Monte, J. Hernando, X. Miró, A. Adolf, "Text independent speaker identification on noisy environements by means of self organizing maps", ICSLP'96, pp1804-1806.

[8] K. R. Farrell, R. J. Mammone, K. T. Assaleh, "Speaker recognition using neural networks and conventional classifiers", IEEE Transactions on speech and audio processing, Vol 2 nº1, part II, pp. 194-205, January 1994.

[9] Y. Linde, A. Buzo, R. M. Gray, "An algorithm for vector quantizer design", IEEE Transactions on Communications, vol. COM-28, pp. 84-95, January 1980.

[10] T. Masters, "Advanced Algorithms for Neural Networks. A C++ Sourcebook", John Wiley & Sons, New York, pp. 47-71.

| Codebook size (bits) | LPCC-VQ | LPCC-VQ Farrell [8] | Combination MSE criterion | Combination MAD criterion |
|---|---|---|---|---|
| 4 | 10% | 10% | 5,79% | 5,79% |
| 5 | 6,84% | 8% | 4,21% | 2,1% |
| 6 | 6,31% | 5% | 3,68% | 2,1% |
| 7 | 3,68% | 4% | 3,16% | 2,1% |

**Table 1.** Identification error rates as function of codebook size for K=2.

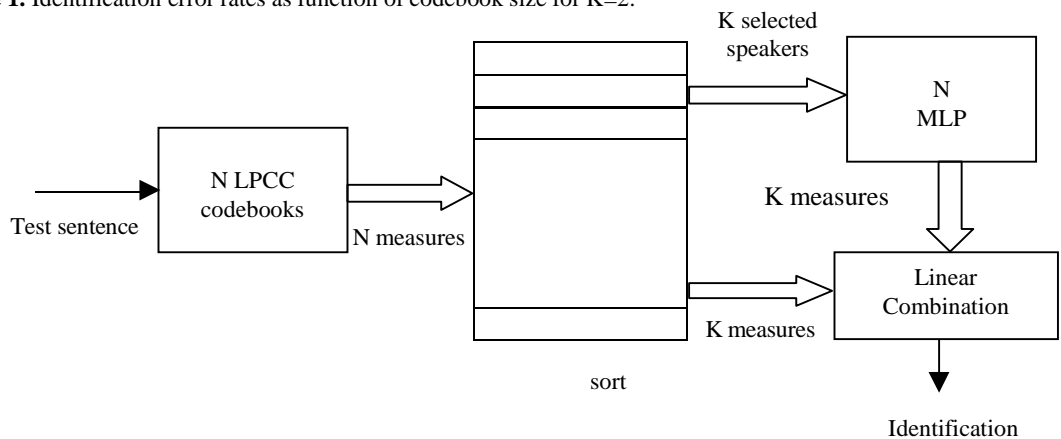

**Figure 1.** Proposed scheme